\documentclass[12pt,a4paper]{article}

 \renewcommand{\baselinestretch}{1.4}

\usepackage{amsmath}
\numberwithin{equation}{section} 

 \newenvironment{myabstract}[1]{
 \renewcommand{\baselinestretch}{1.4}
 \vspace*{10mm} \large
 \begin{center} {\bf Abstract} \end{center}
 \begin{center} \parbox{14cm}{\small #1}}{\end{center}}

\begin{document}

 \renewcommand{\baselinestretch}{1.7}

 \title{\vspace{-20mm} Conservation of energy and Gauss Bonnet gravity}

 \author{Christophe R\'{e}al \\ 22, rue de Pontoise, 75005 PARIS \vspace{-3mm} \\
 \vspace{4mm}}

 \date{November, 2007}

 \maketitle

\vspace{10mm}
 I dedicate this work to Odilia.

 \begin{myabstract}{

In the present article, we prove that any tensor of degree two in
the Riemann tensor, which satisfies the principle of conservation
of energy, vanishes identically, in dimension $D=4$, because it
comes from the topological Gauss-Bonnet lagrangian. We conjecture
that, in dimension $D=2n$, the topological Euler form can also be
deduced from the conservation of energy. We then give arguments in
favor of complex gravity, and write a tensor that is likely to
describe quantum gravity, with a dimensionless gravitational
coupling constant.

 }\end{myabstract}

\vspace{10mm}

\section{Introduction}
\subsection{Some well known facts}
\paragraph{Conservation of energy}
If we look at the Einstein equations $R_{ik}-\frac{1}{2}Rg_{ik}=
\kappa T_{ik}$ of general relativity, we see that the
gravitational part is composed of a tensor $G_{ik}$, which
verifies minimal conditions for the equation possible. The first
condition, which enabled Einstein to find out his tensor, is that
it should be constructed out of second derivatives of the
fundamental variables of the theory, which are the $g_{ik}$.
Mathematically, this means that $G_{ik}$ must be constructed from
the curvature tensor. Looking at the other side of the equation,
we immediately see another necessary condition on $G_{ik}$,
imposed by the law of conservation of energy $\nabla^{i}T_{ik}=0$
on the matter tensor. So the equation is possible if and only if
$\nabla^{i}G_{ik}=0$. In fact, the tensor calculus provides us
with this equation by a formal computation.
\paragraph{Dimension and topology}
However, the tensor $G_{ik}=R_{ik}-\frac{1}{2}Rg_{ik}$ has
dramatically different properties, depending on the dimension of
space-time, and particularly its properties are different in the
case $D=2$ and when $D\geq3$. In dimension $D=2$, the
Hilbert-Einstein action $\int\sqrt{-g}R$ is topological,  and the
Einstein tensor $R_{ik}-\frac{1}{2}Rg_{ik}$ possesses the
condition of conformal invariance, we mean that its trace
vanishes.
\subsection{Constructing other tensors for gravity}
\paragraph{A dimensionless coupling constant}
Using these first remarks, we consider the mathematical problem to
construct all possible tensors $\Sigma_{ik}$, made of the
curvature tensor, and verifying the necessary law of conservation
of energy : $\nabla^{i}\Sigma_{ik}=0$. We will see that, if
$G_{ik}$ is the only tensor made of $R_{ijkl}$, of degree one in
$R_{ijkl}$, and verifying the law of conservation of energy, there
also is a unique tensor made of $R_{ijkl}$, of degree two in
$R_{ijkl}$, and verifying the same law. The essential feature of
this tensor is that possesses, in dimension $D=4$, the properties
of the Einstein tensor in $D=2$. It is conformal invariant, we
mean that its trace vanishes, and it has a dimensionless
gravitational coupling constant. It is clear that it can
conjectured that there exists, for each integer $n$, a unique
tensor made of $R_{ijkl}$, of degree n in $R_{ijkl}$, which is
conformal invariant and which possesses a dimensionless coupling
constant in dimension $D=2n$
\paragraph{When topology appears}
In fact, these tensors of degree $n$ in $R_{ijkl}$ have, in their
respective dimension $D=2n$, another property of Einstein's tensor
in $D=2$ : they are trivial, because they are topological. Thus,
in dimension $D=4$, starting from a tensor of degree $2$ in
$R_{ijkl}$, we can see in the calculation the following striking
property : the sole condition of conservation of energy, makes
appear in our tensor the exact coefficients of the topological
Gauss-Bonnet term. In dimension $D=2n$, the mathematical
conjecture is that the sole equation of conservation of energy
makes appear in the tensor of degree $n$ in $R_{ijkl}$ the
coefficients of the Euler form. Since Donaldson invariants and
then Seiberg-Witten invariants, we know a lot about the relations
between physics and topology. Here, we use such a simple and
direct relation between these two fields to construct another kind
of quantum equation of gravity.
\paragraph{Complex gravity and the other quantum interactions}
In the quantum context, the wavy nature of matter is reflected by
the fact, that in some way, complex field variables come into
play. Looking carefully at a list in which all energy-momentum
tensors, ready to quantization, are written down and put together,
(Grib, Mamayev, Mostepanenko 1992, \cite{GMM} Part I, Chapter 1),
and by simple inspection, we observe general quantum features :
all these tensors are of degree two in complex field variables and
the doubling is made via complex conjugates. Applying the same
rules, by analogy, to gravity, we arrive at a natural conclusion :
gravity should be complex, we mean $g_{ik}$ should be complex, the
tensor for gravity should be of degree $2$, it thus should be the
complex analog of the vanishing topological real tensor of degree
$2$, which makes appear $D=4$ as a preferred dimension of
space-time. We will not investigate more this complex tensor here,
but if, in the complex case, this tensor is effectively non
vanishing, we believe these links between reality and complexity,
conservation of energy and topology, could be the key to
understand why our world possesses four dimensions.
\section{The tensor of degree two}
\subsection{Einstein's tensor of degree one}
We just remember how we prove the existence and the uniqueness of
$G_{ik}$ of degree one in $R_{ijkl}$. As $G_{ik}$ is of degree one
in $R_{ijkl}$, only can it contain $R_{ik}$ and $R$. So we have
$G_{ik}=R_{ik}+\alpha R g_{ik}$ where $\alpha$ is a constant to be
determined. Using the tensor calculus which gives formally
$\nabla^{i}R_{ik}=\frac{1}{2}\partial_{k}R$, we see that
$\nabla^{i}G_{ik}=(\alpha+\frac{1}{2})\partial_{k}R=0$ if and only
if $\alpha=-\frac{1}{2}$. This gives the existence and the
uniqueness of the tensor, as well as its exact expression. We now
study the case of the degree two.
\subsection{Ingredients for the tensor of degree two}
\paragraph{The method} Many more terms will contain the
tensor of degree two, because in this case, there are the
possibilities of using the four indexed $R_{ijkl}$, with indices
contracted, as in $R_{iabc}{R_{k}}^{abc}$ or as in
$R_{iakb}R^{ab}$. So we first have to determine all possible
terms, and then calculate all the constants appearing in the
linear combination forming our tensor. We recall that we note
$\Sigma_{ik}$ for this tensor. Next, using the law of conservation
of energy for $\Sigma_{ik}$, we show that there is a unique
solution to this set of constants.
\paragraph{The general form of the tensor} To find the components
of $\Sigma_{ik}$ of degree two, we have simply to multiply two
tensors of the form $R_{abcd}$, $R_{ab}$, or $R$ and use as well
$g_{ik}$, where the indices $a,b,c,d$ are chosen to be $i$ or $k$,
or are otherwise contracted. \paragraph{Products containing the
scalar curvature $R$} For a product $R^{2}$ the only possible term
is $R^{2}g_{ik}$, for a product of $R_{ab}$ with $R$, again one
possibility, which is $RR_{ik}$.
\paragraph{Products Ricci-Ricci} For two products of $R_{ab}$, the
indices $i$ and $k$ must belong to different $R_{ab}$, to avoid
the appearance of the contraction $R$, a case already studied, and
using that $R_{ab}$ is symmetric, we get the only
$R_{ia}{R_{k}}^{a}$.
\paragraph{Products Ricci-Riemann} For products of $R_{ab}$ and
$R_{abcd}$, the term $R_{ik}$ cannot appear, otherwise the
contraction of $R_{abcd}$ is $R$. As well, if $R_{ia}$ appears,
using the symmetries in the indices of $R_{abcd}$, we can suppose
that $k$ is the first index. We have then an expression of the
form $R_{ia}R_{kpqr}$, where, among the indices, two are in the up
position, one in the down position, $a$ appears once, in the up
position, to be contracted with the index $a$ of $R_{ia}$, and say
$b$ appears twice among $p$, $q$ and $r$, and is contracted. Then,
in this Riemann tensor $a$ cannot be the second index, otherwise
the contraction over $b$ is zero, so we can suppose $a$ is the
third index, and the contraction over $b$ gives us another Ricci.
So nor $i$ neither $k$ can appear in the Ricci, and we have then
an $R^{ab}$ where $a,b$ are to be contracted with indices of a
Riemann tensor. As $R^{ab}$ is symmetric in $a,b$, it cannot be
contracted with indices $a,b$ placed in an antisymmetric position
in $R_{pqcd}$, and as $R_{pqcd}$ is antisymmetric in the first two
indices and also in the last two, there is one $a$ in the first
two and one $b$ in the last two. Using again the symmetries of the
indices in the Riemann tensor, we can chose $i$ in first place and
$k$ in the third, and we are left with the only possibility
$R^{ab}R_{iakb}$.
\paragraph{Products Riemann-Riemann} For the product of two Riemann
tensors, it is quite direct to see that the only possibility is
${R_{i}}^{abc}R_{kabc}$. First, as before, we can suppose that $i$
is the first index of the first Riemann. Now if $i$ and $k$ appear
only in the first Riemann, $c$ for example appears twice in the
second, giving us zero or Ricci. So $k$ is the first index of the
second Riemann. Now we can chose the first as ${R_{i}}^{abc}$ and
using the antisymmetry of the second tensor in the last two
indices, we can suppose that in it, the last two indices are in
alphabetical order. We are left with $R_{kabc}$, $R_{kbac}$ and
$R_{kcab}$. Using now that in the first Riemann $b$ and $c$ appear
in antisymmetric positions, we have
${R_{i}}^{abc}R_{kbac}=-{R_{i}}^{abc}R_{kcab}$. Using finally the
identity $R_{kabc}-R_{kbac}+R_{kcab}=0$, we see that all possible
tensors can be written only in terms of ${R_{i}}^{abc}R_{kabc}$.
\paragraph{Terms involving the metric tensor} In all this, we have
discarded the possibility of the appearance of $g_{ik}$, but the
same arguments permit to conclude that the only possible terms are
$R^{(4)}g_{ik}$ where $R^{(4)}=R^{abcd}R_{abcd}$, $R^{(2)}g_{ik}$
where $R^{(2)}=R^{ab}R_{ab}$ and of course the $R^{2}g_{ik}$ first
considered. \paragraph{Synthesis} To conclude we get then the most
general tensor $\Sigma_{ik}$ of degree two :
\begin{equation}
\Sigma_{ik}={R_{i}}^{abc}R_{kabc}+\alpha R_{iakb}R^{ab}+\beta
R_{ia}{R_{k}}^{a}+\gamma R_{ik}R+\left(\delta R^{(4)}+\epsilon
R^{(2)}+\eta R^{2}\right)g_{ik}
\end{equation}
\subsection{Three formulas}
In order to calculate the coefficients appearing in
$\nabla^{i}\Sigma_{ik}$, we need a first formula :
\begin{equation}
\nabla^{i}R_{iabc}=\nabla_{b}R_{ac}-\nabla_{c}R_{ab}
\end{equation}
Starting with the Bianchi identity :
\begin{equation}
\nabla_{m}R_{nabc}+\nabla_{c}R_{namb}+\nabla_{b}R_{nacm}=0
\end{equation}
and contracting over m and n, we obtain (24.2) directly :
$$\nabla^{n}R_{nabc}+\nabla_{c}R_{ab}-\nabla_{b}R_{ac}=0 $$ Here, we
adopt the convention that the contraction of the first and the
third indices in the Riemann tensor gives the Ricci tensor, and
then the contraction of the first and fourth indices in the
Riemann tensor gives minus the Ricci tensor, because of the
antisymmetry of third and fourth indices in the Riemann tensor.
This gives the formula (2.2). Now, we calculate the coefficients
of $\nabla^{i}\Sigma_{ik}$ one by one. First we need a second
formula :
\begin{equation}
\nabla^{i}({R_{i}}^{abc}R_{kabc})=(\nabla_{b}R_{ac}){R_{k}}^{abc}-(\nabla_{c}R_{ab}){R_{k}}^{abc}+2R^{iabc}(\nabla_{c}R_{kabi})
\end{equation}
Using the properties of the connection $\nabla$, we find :
\begin{equation}
\nabla^{i}({R_{i}}^{abc}R_{kabc})=(\nabla^{i}{R_{i}}^{abc})R_{kabc}+{R_{i}}^{abc}(\nabla^{i}R_{kabc})
\end{equation}
and using
\begin{equation}(\nabla^{i}{R_{i}}^{abc})R_{kabc}=(\nabla^{i}R_{iabc}){R_{k}}^{abc}\end{equation}
as well as equation (2.2), we obtain immediately that the first
term of the right hand side of (2.5) equals the first two terms of
formula (2.2). So, we only need to prove that the second term on
the right hand side of (2.5) equals the third of formula (2.4).
Now, using the Bianchi identity (2.3), we have :
\begin{equation}{R_{i}}^{abc}(\nabla^{i}R_{kabc})=-{R_{i}}^{abc}\nabla_{b}{R_{kac}}^{i}-R^{iabc}\nabla_{c}R_{kaib}\end{equation}
Using the antisymmetry of the indices $b$ and $c$ in the first
Riemann tensor of the term
$-{R_{i}}^{abc}\nabla_{b}{R_{kac}}^{i}$, we obtain that this term
equals ${R_{i}}^{abc}\nabla_{c}{R_{kab}}^{i}$, which also equals
the term $-R^{iabc}\nabla_{c}R_{kaib}$, using the antisymmetry of
$i$ and $b$ in the second Riemann tensor. Altogether, we see that
formula (2.4) has been proved. Now, we study the term arising in
$\nabla^{i}\Sigma_{ik}$ from the second term of $\Sigma_{ik}$,
asserting the following formula :
\begin{equation}
\nabla^{i}(\alpha R_{iakb}R^{ab})=\alpha
R^{ab}(\nabla_{k}R_{ab})-\alpha R^{ab}(\nabla_{b}R_{ak})-\alpha
(\nabla_{c}R_{ab}){R_{k}}^{abc}
\end{equation} To prove it, we use :
$$\nabla^{i}(R_{iakb}R^{ab})=(\nabla^{i}R_{iakb})R^{ab}+R_{iakb}\nabla^{i}R^{ab}$$
\begin{equation}
=(\nabla^{i}R_{iakb})R^{ab}+{{R^{ia}}_{k}}^{b}\nabla_{i}R_{ab}
\end{equation}
Now, applying to the first term on the right hand side of (24.9)
the identity (2.2), we readily obtain the first two terms on the
right hand side of formula (2.8). But the second term on the right
hand side of (2.9) can be written
$(\nabla_{c}R_{ab}){{R^{ca}}_{k}}^{b}$ and since exchanging the
two pairs of indices in the Riemann tensor does not change its
value, we obtain $(\nabla_{c}R_{ab}){R_{k}}^{bca}$. Now, using the
well known identity
\begin{equation}
R_{iklm}+R_{imkl}+R_{ilmk}=0
\end{equation}
this term becomes
$$-(\nabla_{c}R_{ab}){R_{k}}^{abc}-(\nabla_{c}R_{ab}){R_{k}}^{cab}$$
In this last equation, the second term gives zero because $a,b$
are contracted and appear in symmetric positions in the Ricci
tensor and in antisymmetric positions in the Riemann tensor. This
finishes the proof of formula (2.8).
\subsection{Computing the coefficients of the tensor}
Taking the $\beta$ and $\gamma$ terms of $\Sigma_{ik}$, and
remembering that $\nabla^{i}R_{ik}=\frac{1}{2}\partial_{k}R$, we
find at once :
\begin{equation}
\nabla^{i}(\beta R_{ia}{R_{k}}^{a})=\frac{1}{2} \beta
(\partial^{a}R)R_{ak}+ \beta R^{ia} \nabla_{i} R_{ka}
\end{equation}
and
\begin{equation}
\nabla^{i}(\gamma R_{ik}R)=\frac{1}{2} \gamma (\partial_{k}R)R+
\gamma R_{ik} (\partial^{i}R)
\end{equation}
\paragraph{Computing $\alpha$, $\beta$, $\gamma$}
Looking closely at our equations, we see that the second terms of
(2.8) and (2.11) can be eliminated by the choice $\alpha=\beta$,
and that the first term of (2.11) gives zero, when combined with
the second term of (2.12), provided that we choose the relation
$\beta=-2\gamma$. So, by simple inspection of our equations, we
possess an easy way to calculate our coefficients.
\paragraph{A relation which simplifies the whole calculation}
Turning now to the $\delta$-term, we have :
$$\nabla^{i}\left(R^{(4)}g_{ik}\right)= \nabla^{i}[R_{abcd} R^{abcd}g_{ik}] = 2 (\nabla_{k}R_{abcd}) R^{abcd}$$
which equals :
$$-2R^{abcd}\nabla_{d}R_{abkc}-2R^{abcd}\nabla_{c}R_{abdk}$$
because of (2.3). Both of these terms equal
$2R^{abcd}\nabla_{c}R_{abkd}$, the second because in the second
Riemann tensor, $d$ and $k$ are in antisymmetric positions, and
the first because in the first Riemann tensor, $c$ and $d$ are in
antisymmetric positions. We thus find :
$$\delta\nabla^{i}\left(R^{(4)}g_{ik}\right)= \delta \nabla^{i}[R_{abcd} R^{abcd}g_{ik}]= 4 \delta R^{abcd}\nabla_{c}R_{abkd}$$
$$ =4 \delta \nabla_{c}(R^{abcd}R_{abkd}) -4 \delta R_{abkd}\nabla_{c}R^{abcd}$$
Now happens a considerable simplification, because the first term
of the former equation can be written $4 \delta
\nabla^{c}(R_{abcd}{{R^{ab}}_{k}}^{d})$. Thus, $c$, which is
contracted, can be called $i$, and we can exchange the two pairs
of indices in both Riemann tensors, obtaining : $4 \delta
\nabla^{i}(R_{idab}{R_{k}}^{dab})$. This term is exactly the term
of $\nabla^{i}\Sigma_{ik}$ which corresponds to the first term in
the sum giving $\Sigma_{ik}$. So we find that choosing $\delta = -
\frac{1}{4}$, we eliminate all the terms of (2.4). We are now left
with a very few terms, the first and the third terms on the right
hand side of (2.8), the first term on the right hand side of
(2.12), the last $-4 \delta R_{abkd}\nabla_{c}R^{abcd}$ and
finally the $\epsilon$ and $\eta$ terms of (2.1).
\paragraph{Computation of the $\delta$-term} This term can be
written
$$-4 \delta (\nabla^{c}R_{abcd}){{R^{ab}}_{k}}^{d}= (\nabla_{a}R_{db}-\nabla_{b}R_{ad}){{R^{ab}}_{k}}^{d}$$
using the value of $\delta$ and also formula (2.2). The second
term on the right hand side of this formula is equal to the first,
because in the Riemann tensor, $a$ and $b$ appear in antisymmetric
positions, and we are left with :
$$2(\nabla_{a}R_{bd}){{R^{ab}}_{k}}^{d}=2(\nabla_{a}R_{bd}){R_{k}}^{dab}=2(\nabla_{c}R_{ab}){R_{k}}^{bca}$$
Indeed, we obtain the first equality by exchanging the pairs of
indices in the Riemann tensor, and the second by renaming
contracted indices. We use formula (2.10) to write
${R_{k}}^{bca}=-{R_{k}}^{abc}-{R_{k}}^{cab}$, and we observe that
the second term has $a$ and $b$ in antisymmetric positions, which
gives zero because these indices are contracted with
$\nabla_{c}R_{ab}$. So the calculation of the $\delta$-term of
$\nabla^{i}\Sigma_{ik}$ is finished, and gives us only
$-2(\nabla_{c}R_{ab}){R_{k}}^{abc}$, this term vanishing with the
third term of (2.8) if and only if $\alpha = -2$. Comparing this
result with the other relations obtained for $\beta$ and $\gamma$,
we now find $\beta=-2$, and $\gamma=+1$.
\paragraph{The $\epsilon$-term} We are now ready to study the
$\epsilon$-term. We know that we still have to cancel the first
term of (2.8) and the first term of (2.12).
$$\epsilon \nabla^{i}[R_{ab} R^{ab}g_{ik}] = 2 \epsilon
(\nabla_{k}R_{ab}) R^{ab}$$ cancels directly the first term of
(2.8) provided $2\epsilon=-\alpha$, so $\epsilon=+1$.
\paragraph{The $\eta$-term} The $\eta$-term gives $\eta
\nabla^{i}[R^{2}g_{ik}]=2\eta R(\partial_{k}R)$, cancelling
 the first term of (2.12) provided $2\eta=-\frac{1}{2}\gamma$, which
leads to $\eta = - \frac{1}{4}$, providing us finally a set of
constants for which $\nabla^{i}\Sigma_{ik}=0$. Finally we proved
the statement of existence in the following theorem :
\paragraph{Theorem :}There exists a unique tensor $\Sigma_{ik}$,
constructed from all possible products of degree two of the
Riemann tensor, its contractions, and the metric tensor, which
verifies the law of conservation of energy :
$\nabla^{i}\Sigma_{ik}=0$. This tensor contains effectively all
possible products and has the form :
\begin{equation}
\Sigma_{ik}={R_{i}}^{abc}R_{kabc}- 2 R_{iakb}R^{ab} - 2
R_{ia}{R_{k}}^{a} + R_{ik}R - \frac{1}{4}\left( R^{(4)} -4
R^{(2)}+ R^{2}\right)g_{ik}
\end{equation}
where $R^{(4)}=R^{abcd}R_{abcd}$ and $R^{(2)}=R^{ab}R_{ab}$
\paragraph{Existence} As we said, the existence in the theorem has
been proved before, we just notice that we used for this proof all
identities we know concerning the Riemann tensor and its
contractions.
\paragraph{Uniqueness}Of course, we have also proved that there was
no more possible products which could be ingredients of the tensor
$\Sigma_{ik}$, and that our coefficients formed the complete set
of degrees of freedom of our mathematical problem. Finding these
coefficients has been possible because we could cancel all terms
in $\nabla^{i}\Sigma_{ik}=0$, using the well known relations on
the Riemann tensor. It appears that, as there does not exist any
such other relation on this tensor, available in the generic
situation, this was the unique manner of cancelling these terms,
and that the coefficients of $\Sigma_{ik}$ are unique. Here, we
give a method to obtain an explicit proof of the uniqueness of
$\Sigma_{ik}$ : starting with the value of $\Sigma_{ik}$ with all
its coefficients, at first undetermined, we compute
$\nabla^{i}\Sigma_{ik}$ in different explicit choices of the
metric $g_{ik}$, and each example gives us a linear combination of
our coefficients, that we put equal to zero. So we find a linear
system for these coefficients and with enough choices of different
$g_{ik}$, we obtain enough equations, to prove finally that only
the coefficients of the theorem give zero in the generic
situation.
\section{Higher dimensions, topology and complex gravity}
\subsection{The conjecture in higher dimensions}
From what has been done in the case of degree two, it is easily
guessed what can be done as well in the case of degree $n$. We can
consider a tensor $\Sigma_{ik}$, of degree $n$ in the Riemann
tensor and its contractions, and first find all possible products
of degree $n$ that could appear in $\Sigma_{ik}$. Then, we find
all coefficients by imposing that in $\nabla^{i}\Sigma_{ik}$, all
terms vanish. Looking at the case $n=1$ and $n=2$, it should be
clear that it is a way of proving the following conjecture :
\paragraph{Conjecture :}There is a unique tensor $\Sigma_{ik}$
constructed from all possible products of degree $n$ in the
Riemann tensor and its contractions, constructed with the metric
tensor too, and which verifies the law of conservation of energy :
$\nabla^{i}\Sigma_{ik}=0$. This tensor has the form :$$\Sigma_{ik}
= {\tilde{\Sigma}}_{ik}-\frac{1}{2n}{\tilde{\Sigma}}g_{ik}$$ where
$\tilde{\Sigma}$ is the Euler form in dimension $2n$, as well as
the trace of ${\tilde{\Sigma}}_{ik}$. Furthermore, $\Sigma_{ik}$
vanishes, becomes it comes, using the calculus of variation, from
the topological Euler lagrangian.
\subsection{Topology} The appearance in the tensor of degree $2$ of the
Gauss-Bonnet term
$$\tilde{\Sigma} = R^{(4)} -4 R^{(2)}+ R^{2}$$
authorizes us to conjecture that our tensor completely vanishes in
dimensions four, because it comes from the topological
Gauss-Bonnet action :
$$\int \sqrt{-g}(R^{(4)} -4 R^{(2)}+R^{2})$$
In dimensions different from four, the same tensor, of degree two,
comes from the would-be-a-Gauss-Bonnet action :
$$\int \sqrt{-g}(R^{(4)} -4 R^{(2)}+ R^{2})$$
We thus have found an interesting method to write, from an a
priori trivial topological action, a non trivial equation : start
from the topological action in dimension $n$, go to another
dimension where the same action is not trivial anymore, and use
the calculus of variation to extract the tensorial equation. Then,
take the tensor, and go back to the critical dimension. The
question is : does the tensor obtained in this way should be
discarded as being trivial or is it relevant to describe some kind
of physics? This has been the first route which we used to find
our equation of quantum gravity. Because the gravitational tensor
of degree $2$ first displays a dimensionless coupling constant,
and second fits so well with the energy-momentum tensors of the
other interactions, even if it is identically zero, we though
there should be some kind of physics behind. We finally retained
the idea of keeping only its trace in the equation, which gave the
$\Lambda$ term, the law of conservation of energy being in the
equation of quantum gravity being provided by the variations of
$\kappa(\epsilon)$.
\paragraph{A dimensionless coupling constant}
Forgetting that our tensor vanishes for one moment, we consider
the equation that such a tensor would give :
$$\Sigma_{ik} = \kappa T_{ik}$$ To determine the dimension of the
coupling constant, we look at :
$$\Sigma_{0}^{0} = \kappa T_{0}^{0}$$
Here the Riemann and Ricci tensors, when containing the same
number of up and down indices, as well as the scalar curvature,
have dimension $[L]^{-2}$, where $[L]$ is a length. So,
$\Sigma_{0}^{0}$ has dimension $[L]^{-4}$. Now, $T_{0}^{0}$ equals
$\epsilon$, the energy density of matter, and has dimension, in
dimension $D=4$, $[E][L]^{-3}\sim\hbar[T]^{-1}[L]^{-3}\sim\hbar
c[L]^{-4}$, since energy $[E]$ has dimension $\hbar[T]^{-1}$ and
where of course $[T]$ is a time. Comparing these two results, we
see that
$$ \kappa = \frac{\kappa_{0}}{\bar{h}c}$$
where $\kappa_{0}$ has no dimension at all.
 \subsection{Complex gravity} We know that our tensor $\Sigma_{ik}$
probably vanishes because it is the energy-momentum tensor coming
from a topological lagrangian by the calculus of variations, but
there is another form of this tensor, which at least at first
sight, is not necessarily trivial, and which could prove itself
very interesting. Because it is of second order in the curvature
tensor, $\Sigma_{ik}$ possesses a natural extension to complex
gravity. As in the quantum tensors describing particles of
different spin, we can write down a tensor of degree two by
doubling the curvature terms by complex conjugates. By inspection
of these known quantum tensors, we guess easily the procedure to
follow. Indeed, we pose as new fundamental variables, the complex
metric $g_{ik}$ verifying the condition :
\begin{equation}
 g^{*}_{ki}=g_{ik}
\end{equation}
where $z^{*}$ corresponds to the complex conjugate of $z$, and we
pose the complex tensor :
$$\Sigma_{ik}=\frac{1}{2}{{R^{*}}_{i}}^{abc}R_{kabc}+\frac{1}{2}{{R}_{i}}^{abc}R^{*}_{kabc}-{R^{*}}_{iakb}R^{ab}-
R_{iakb}{R^{*}}^{ab}-{R^{*}}_{ia}{R_{k}}^{a}-{R}_{ia}{R^{*}_{k}}^{a}$$
\begin{equation}
+\frac{1}{2}{R^{*}}_{ik}R +\frac{1}{2}R_{ik}{R^{*}}
 - \frac{1}{4}\left(R^{(4)} -4 R^{(2)}+ R{R^{*}}\right)g_{ik}
\end{equation}
where $R^{(4)}={R^{*}}^{abcd}R_{abcd}$ and $R^{(2)} =
{R^{*}}^{ab}R_{ab}$. Equations (3.1) and (3.2) should normally
imply that $\Sigma_{ik}$ is a real symmetric tensor which verifies
the condition of conservation of energy.

\vspace{10mm} Email address : cristobal.real@hotmail.fr

\end{document}